\documentclass{ws-procs9x6}

\usepackage{natbib}

\usepackage{graphicx}

\usepackage{amssymb}

\def\aj{AJ }%
\def\araa{ARA\&A }%
\def\apj{ApJ }%
\def\apjl{ApJ }%
\def\apjs{ApJS }%
\def\aap{A\&A }%
\def\mnras{MNRAS }%
\def\nat{Nature }%
\def\iaucirc{IAU~Circ. }%
\begin{document}

\title{Jets in Supermassive and Stellar-Mass Black Holes}

\author{I.F. Mirabel\footnote{e-mail: {\sf fmirabel@cea.fr};
\uppercase{URL:} {\sf http://www.iafe.uba.ar/astronomia/mirabel/mirabel.html} }}

\address{Centre d'Etudes de Saclay/CEA/DSM/DAPNIA/SAP 91911
Gif/Yvette, France \& Intituto de \protect{Astronomía} y
\protect{Física} del Espacio/CONICET, Argentina}

\maketitle

\abstracts{ 
Relativistic outflows are a common phenomenon in accreting black
holes. Despite the enormous differences in scale, stellar-mass black
holes in X-ray binaries and collapsars, and super-massive black holes
at the dynamic centre of galaxies are sources of jets with analogous
physical properties. Synergism between the research on microquasars,
gamma-ray bursts, and Active Galactic Nuclei should help to gain
insight into the physics of relativistic jets seen everywhere in the
Universe. 
}

\section{The quasar-microquasar analogy}

\begin{figure}[hhhb]
{\centering
\resizebox*{1.0\textwidth}{!}{\rotatebox{0}{\includegraphics{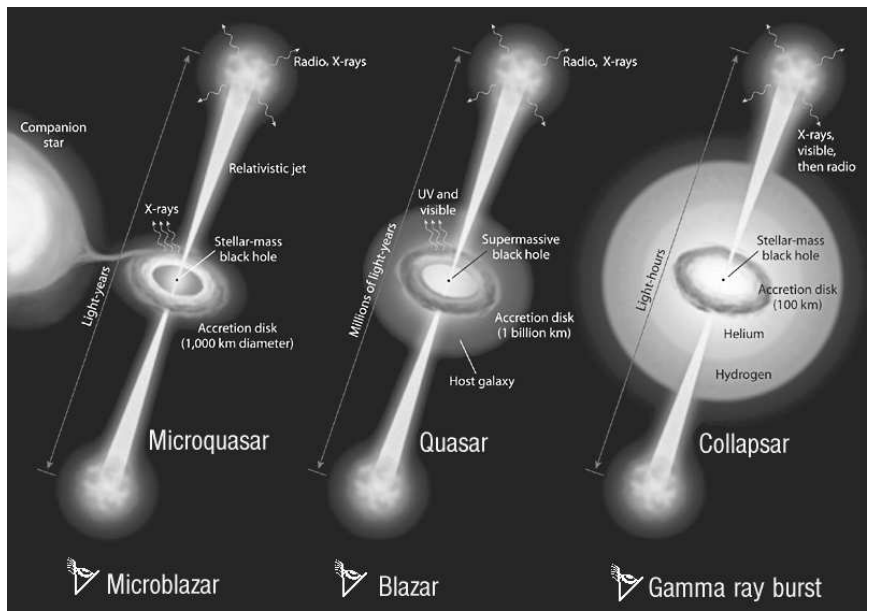}}}
\par}
\caption{Diagram illustrating current ideas concerning microquasars, quasars 
and gamma-ray bursts (not to scale). It is proposed that a universal
mechanism may be at work in all sources of relativistic jets in the
universe. Synergism between these three areas of research in
astrophysics should help to gain a more comprehensive understanding of
the relativistic jet phenomena observed everywhere in the universe.}
\label{fig-1}
\end{figure}

Microquasars are scaled-down versions of quasars and both are believed
to be powered by spinning black holes with masses of up to a few tens
that of the Sun (see Figure~\ref{fig-1}). The word {\it microquasar}
was chosen by \cite{Mirabelnat92} to suggest that we could learn
about microquasars from previous decades of studies on Active Galactic 
Nuclei (AGN). A major difference is that the linear and time
scales of the phenomena are proportional to the black hole mass.
In quasars and microquasars are found the following three basic
ingredients: 1) a spinning black hole, 2) an accretion disk heated by
viscous dissipation, and 3) collimated jets of relativistic
particles. 

Because of the relative proximity and shorter time scales, in
microquasars it is possible to firmly establish the relativistic
motion of the sources of radiation, and to better study the physics of
accretion flows and jet formation near the horizon of black
holes. Jets in microquasars are easier to follow because their
apparent motions in the sky are $\geq$10$^3$ faster than in
quasars. Because microquasars are found in our Galaxy the two-sided
moving jets are more easily seen than in AGN
\cite{Mirabelnat94}. However, to know how the jets are collimated in
units of length of the black hole's horizon, AGN up to distances of a
few Mpc may present an advantage. \cite{Biretta} find that the
initial collimation of the non-thermal jet in the galaxy M87 of the
Virgo cluster takes place on a scale of 30-100 R${_S}$, which is
consistent with poloidal collimation by an accretion disk.

\section{The microquasar gamma-ray-burst analogy}

There is increasing evidence that the central engine of the most
common form of gamma-ray bursts (GRBs), those that last longer than a
few seconds, are afterglows from ultra-relativistic jets produced
during the formation of stellar-mass black holes
\cite{MacFadyen}. \cite{Mirabelrev} proposed that ultra-relativistic
bulk motion and beaming are needed to explain: 1) the enormous energy
requirements of $\geq$ 10$^{54}$ erg if the emission were isotropic
\cite[e.g.][]{Kulkarni,Castro-Tirado}; 2) the
statistical correlation between time variability and brightness
\cite{Ramirez-Ruiz}, and 3) the statistical
anti-correlation between brightness and time-lag between hard and soft
components \cite{Norris}. Beaming reduces the energy release by the
beaming factor f = $\Delta$$\Omega$/4$\pi$, where $\Delta$$\Omega$ is
the solid angle of the beamed emission. Additionally, the photon
energies can be boosted to higher values.  Extreme flows from
collapsars with bulk Lorentz factors $>$ 100 have been proposed as
sources of $\gamma$-ray bursts \cite{Meszaros}. High collimation
\cite{Dado,Pugliese} can be tested observationally \cite{Rhoads},
since the statistical properties of the bursts will depend on the
viewing angle relative to the jet axis.

Recent multi-wavelength studies of gamma-ray afterglows suggest that
they are highly collimated jets. The brightness of the optical
transient associated to some GRBs show a break
\cite[e.g.][]{Kulkarni}, and a steepening from a power law in time t
proportional to t$^{-1.2}$, ultimately approaching a slope t$^{-2.5}$
\cite[e.g.][]{Castro-Tirado}. The achromatic steepening of the
optical light curve and early radio flux decay of some GRBs are
inconsistent with simple spherical expansion, and well fit by jet
evolution. It is interesting that the power laws that describe the
light curves of the ejecta in microquasars show similar breaks and
steepening of the radio flux density \cite{Mirabelrev}. In
microquasars, these breaks and steepenings have been interpreted
\cite{Hjellming88} as 
a transition from slow intrinsic expansion followed by free expansion
in two dimensions. Besides, linear polarizations of about 2\% were
recently measured in the optical afterglows \cite[e.g.][]{Covino},
providing strong evidence that the afterglow radiation from gamma-ray
bursts is, at least in part, produced by synchrotron processes.
Linear polarizations in the range of 2-10\% have been measured in
microquasars at radio \cite[e.g.][]{Rodriguez95}, and optical
\cite{Scaltriti} wavelengths.

The jets in microquasars of our own Galaxy seem to be less extreme
local analogs of the super-relativistic jets associated to the more
distant gamma-ray bursts. But the latter do not repeat, seem to be
related to catastrophic events, and have much larger super-Eddington
luminosities. According to the latest models, the same symbiotic
disk-jet relationship as in microquasars and quasars powers the
GRBs. In fact, it is now believed that the Lorentz factors at the base
of the jets inside the collapsing star are $\leq$10 as in microquasars
and quasars, and they reach values $\geq$100 when they break free from
the infalling outer layers of the progenitor star. Because of the
enormous difference in power, the scaling laws in terms of the black
hole mass that are valid for the analogy between microquasars and
quasars may not apply in the case of gamma-ray bursts.

\begin{figure}[hhht]
{\centering
\resizebox*{1.0\textwidth}{!}{\rotatebox{0}{\includegraphics{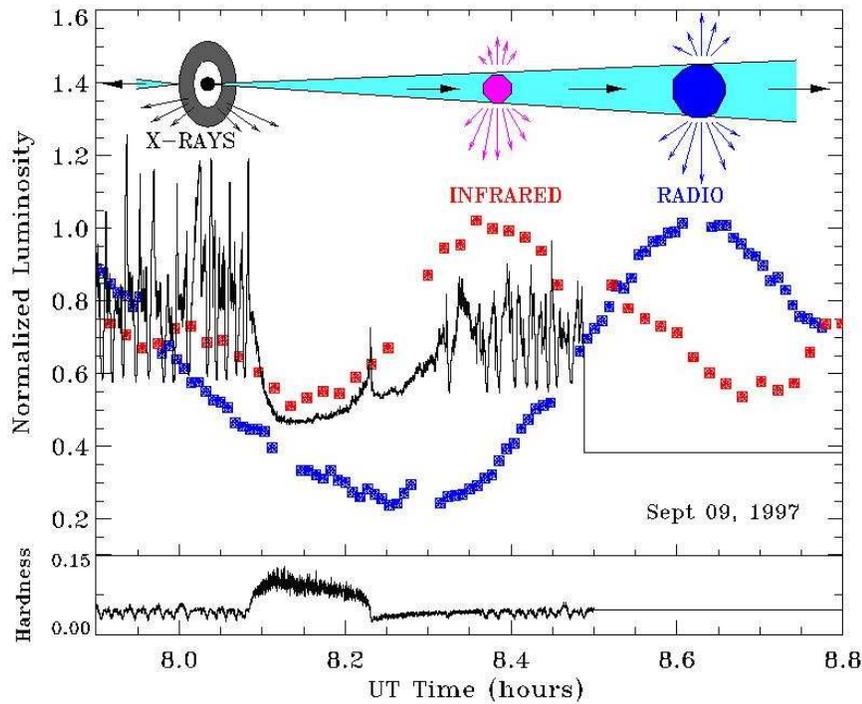}}}
\par}
\caption{Radio, infrared, and X-ray light curves for GRS~1915+105 at
the time of quasi-periodic oscillations with scales of time of
$\sim$20 min \cite{Mirabel98}.  The infrared flare starts during the
recovery from the X-ray dip, when a sharp, isolated X-ray spike-like
feature is observed.  These observations show the connection between
the rapid disappearance and follow-up replenishment of the inner
accretion disk seen in the X-rays \cite{Belloni}, with the ejection of
relativistic plasma clouds observed first as synchrotron emission at
infrared wavelengths, later at radio wavelengths.  A scheme of the
relative positions where the different emissions originate is shown in
the top part of the figure.  The hardness ratio (13-60 keV)/(2-13 keV)
is shown at the bottom of the figure. Analogous phenomena have now
been obseved in the quasar 3C 120 but in time scales of years
\cite{Marscher}}
\label{fig-2}
\end{figure}

\section{Accretion disk origin of relativistic jets}

Synergism between results from multiwavelength simultaneous
observations in microquasars and quasars is providing important
insights into the connection between accretion disk instabilities and
the genesis of jets. Since the characteristic times in the flow of
matter onto a black hole are proportional to its mass, the
accretion-ejection phenomena in quasars should last 10$^5$-10$^7$
longer than analogous phenomena in microquasars \cite{Sams}.
Therefore, variations on scales of tens of minutes of duration in
microquasars could be sampling phenomena that had been difficult to
observe in quasars.

Simultaneous multiwavelength observations of a microquasar revealed in
an interval of time of a few tens of minutes the connection between
the sudden disappearance of the inner $\sim$200 km of the accretion
disk with the ejection of expanding clouds of relativistic plasma (see
Figure~\ref{fig-2}). One possible interpretation of the observations
shown in Figure~\ref{fig-2} is that the plasma of the inner disk that
radiates in the X-rays falls beyond the horizon of the black hole in
$\sim$5min, and subsequently the inner accretion disk is refilled in
$\sim$20 min. While the inner disk is being replenished, we observe
the ejection of a relativistic plasma cloud, first at 2$\mu$m, and
latter at radio wavelengths as the cloud expands and becomes
transparent for its proper radiation at longer wavelengths. The delay
between the maxima at radio and infrared wavelengths is equal to the
one computed with the model for a spherically symmetric expanding
clouds in relativistic AGN jets by \cite{vanderLaan}. Although VLBA
images of these transient ejecta by \cite{Dhawan1} have shown that
they are in fact connical jets, the model first developed for AGN is a
good first approximation, and allows to demonstrate that the infared
flares that preceed the radio flares are synchrotron, rather than
thermal emission. This implies the presence in the jets of electrons
with Lorentz factors $\geq$ 10$^3$ \cite{Fender98,Mirabel98}.

Analogous accretion disk-jet connections were observed in the quasar
3C 120 by \cite{Marscher}. Jets were detected with VLBI after
sudden X-ray dips observed with RXTE, but on scales of a few
years. The scales of time of the phenomena are within a factor of 10
the black hole mass ratios between the quasar and microquasar, which
is relativelly small when compared with the uncertainties in the data.

\section{Compact jets in accreting black holes}

The class of stellar-mass black holes that are persistent X-ray
sources (e.g. Cygnus X-1, 1E 1740-2942, GRS 1758-258, etc.) and some
super-massive black holes at the centre of galaxies (e.g. Sgr A$^*$
and many AGN) do not exhibit luminous outbursts with large-scale
sporadic ejections. However, despite the enormous differences in mass,
steadily accreting black holes have analogous radio cores with steady,
flat (S$_{\nu}$$\propto$$\nu$$^{\alpha}$; $\alpha$$\sim$0) emission at
radio wavelengths. The fluxes of the core component in AGN are
typically of a few Janskys (e.g. Sgr A$^*$$\sim$1Jy) allowing VLBI
high resolution studies, but in stellar mass black holes the cores are
much fainter, typically of a few mJy, which makes difficult high
resolution observations of the core.

From the spectral shape it was proposed that the steady compact radio
emission in black hole X-ray binaries are jets
\cite[e.g.][]{Rodriguez95,Fender99,Fender2000,Corbel}. 
Recently, this has been confirmed by VLBI observations at AU scale
resolution of GRS 1915+105 \cite{Dhawan1}, and Cyg X-1
\cite{Stirling} in the low-hard X-ray state. VLBA images of GRS
1915+105 show compact jets with sizes $\sim$10$\lambda$$_{cm}$ AU
along the same position angle as the superluminal large-scale jets. As
in the radio cores of AGN, the brightness temperature of the compact
jet in GRS 1915+105 is T$_B$$\geq$10$^9$ K. The VLBA images of GRS
1915+105 are consistent with the conventional model of a conical
expanding jet with synchrotron emission \cite{Hjellming88,Falcke99}
in an optically thick region of solar system size. These compact jets
are also found in neutron star X-ray binaries such as LS 5039 (Paredes
et al. 2000) and Sco X-1 \cite{Fomalont}, and are currently
used to track the path of black holes and neutron stars in our Galaxy
\cite[see][for a review]{Mirabelcargese}.

\section{Interaction of jets with the interstellar medium}

If a compact source injects relativistic plasma into its environment,
it is expected that some fraction of the injected power will be
dissipated by shocks, where reacceleration of particles may take
place. Evidences of such interactions are the radio lobes of 1E
1740.7-2942 \cite{Mirabelnat92}, GRS 1758-258 \cite{Rodriguez92}, and
the two lateral extensions in the nebula W50 that hosts at its center
SS 433. The interaction of SS 433 with the shells of W50 has been
studied in the X-rays \cite{Brinkmann96} and radio wavelengths
\cite[][and references therein]{Dubner}.

Besides the well known relativistic jets seen at sub-arcsec scales in
the radio, large-scale jets become visible in the X-rays at distances
$\sim$ 30 arcmin ($\sim$ 25 pc) from the compact source
\cite{Brinkmann96}. In the radio and X-rays, the lobes reach distances
of up to 1$^\circ$ ($\sim$50 pc).  These large-scale X-ray jets and
radio lobes are the result of the interaction of the mass outflow with
the interstellar medium. From optical and X-ray emission lines it is
found that the sub-arcsec relativistic jets have a kinetic energy of
$\sim$ 10$^{39}$ erg s$^{-1}$ \cite{Margon84}, which is several orders
of magnitude larger than the energy radiated in the X-rays and in the
radio.  \cite{Dubner} estimate that the kinetic energy transferred
into the ambient medium is $\sim$ 2 10$^{51}$ ergs, thus confirming
that the relativistic jets from SS433 represent an important
contribution to the overall energy budget of the surrounding nebula
W50.

Large-scale, decelerating relativistic jets from the microquasar XTE
J1550-564 have been discovered with CHANDRA by \cite{Corbel2002}. The
broadband spectrum of the jets is consistent with synchrotron emission
from electrons with Lorentz factors of $\sim$10$^7$ that are probably
accelerated in the shock waves formed by the interaction of the jets
with the interstellar medium. \cite{Corbel2002} demonstrated that in
microquasars we can study in real time the formation and dynamical
evolution of the working surfaces (lobes) of relativistic jets far
away from the centres of ejection, on time scales inaccessible for
AGN. Working surfaces of microquasar jets as in SS433 and XTE
J1550-564 are potential sources of cosmic rays.

\section*{Acknowledgments}
I thank Irapuan Rodrigues for help with the original manuscript.




\end{document}